\documentclass[aps,pre,reprint,showpacs,showkeys,noeprint,superscriptaddress]{revtex4-1} 
\usepackage[utf8]{inputenc}
\usepackage[T1]{fontenc}
\usepackage{amsmath,amssymb,mathtools}
\usepackage{txfonts}

\usepackage{mathrsfs}

\usepackage{graphicx,grffile}
\graphicspath{{/}}

\usepackage{textcomp}
\usepackage{booktabs,tabularx,dcolumn}
\newcolumntype{d}[1]{D{.}{.}{#1}}

\usepackage{url,hyperref}
\usepackage[usenames,dvipsnames]{xcolor}
\hypersetup{colorlinks=true, linkcolor=BrickRed, urlcolor=blue!50!black, citecolor=blue!50!black}

\usepackage[capitalize]{cleveref}
\crefrangelabelformat{equation}{(#3#1#4)$-$(#5#2#6)}

\renewcommand\rho\varrho
\renewcommand\vec[1]{\textrm{\bfseries #1}}

\usepackage[normalem]{ulem}

\begin{document}
\title{Moving Cooling Source Induced Phase Separation in Binary Liquids: an interplay of competing velocities}

\author{Lakshmipriya K, Harssh Karn, and Sutapa Roy}
\email{sutapa.roy@hyderabad.bits-pilani.ac.in}
\affiliation{Department of Physics, Birla Institute of Technology and Science, Pilani, Hyderabad Campus, Jawahar Nagar, Kapra Mandal, Medchal District, Telangana 500078, India}

\date{\today}

\begin{abstract}
We investigate phase separation dynamics in a binary mixture subjected to a moving cooling source from which cold temperature fronts propagate radially outward into the mixture. 
The motion of the source introduces two distinct velocity scales: 
$v_s$ associated with the translation of the source, and $v$ related to the 
propagation of the cooling thermal fronts. 
Competition between the two velocities determines how long a region of the fluid experiences a temperature change, which directly controls phase separation.
A modified Cahn–Hilliard–Cook framework is employed, incorporating explicit coupling between the time-dependent temperature and concentration fields. 
Our numerical simulation results reveal that the evolving patterns and kinetics 
strongly depend on both the ratio and absolute magnitudes of these two competiting velocities. 
Same value of $v_s/v$ yields distinctly different patterns for different $v$.
The temperature profile delineating spatial regions with local temperatures above and below the demixing temperature controls the shape of the patterns formed. 
The rich parameter space enables one to engineer desired pattern structures by tuning the two velocities. 
\end{abstract}

\maketitle

\section{Introduction}
The phenomenon of phase separation continues to attract substantial 
scientific interest owing to its relevance across a broad spectrum of 
systems and technologies. 
While conventional studies have largely focused on isothermal conditions \cite{puri2009kinetics, bray2002theory, onuki2002phase}, 
phase separation under the influence of temperature gradients 
represents a particularly intriguing yet under-explored domain. 
Such gradients play a pivotal role in a variety of physical, chemical, and 
biological systems. 
Spatial variations in temperature can drive concentration redistribution
and thermophoretic motion \cite{duhr2006, amaya2025}. 
Recently, laser induced phase separation has attracted growing research attention 
due to its ability to provide localized and controllable energy source 
\cite{zhou2025, ktafi2024, zhou2025}. 
Lasers can locally heat a polymer mixture above its lower critical temperature, 
triggering phase separation in selected `local' regions. 
Optical heating can locally modulate temperature and enhance/suppress 
biomolecular condensates \cite{condensate2025}. 
This is highly relevant for membrane-less organelles, stress granules and 
neurodegenartive diseases. 
By tuning temperature gradients, one can engineer anisotropic polymer blends, 
which offer major improvement in permeability and cleanability 
as compared to their isotropic counterparts \cite{voit2005, Hong_2010}.
However, the transient morphological evolution underlying these 
anisotropic structures is still poorly understood. 
It also holds applications in perovskite solar cells where 
it can create ion migration and segregation and thus influencing efficiency.
Recently, thermally regenerative flow batteries have garnered attention 
as sustainable energy sources, leveraging liquid-liquid phase separation 
in response to imposed thermal gradients - commonly referred to
as thermo-responsive phase separation \cite{matsui2023}. 
Despite its technological significance, theoretical understanding
of thermo-responsive phase separation remains limited.

A comprehensive understanding of such systems requires accounting for 
both temporally and spatially varying temperature and composition fields, 
along with a coupling between them. 
The composition would be the concentration and density field for a binary mixture and single-component liquid, respectively. 
This coupling significantly increases the mathematical complexity, 
thus making full analytical solution often intractable.
Consequently, numerical studies and simulations are essential in gaining insights. 
The dynamic critical behavior of solids were studied by exposing Ising ferromagnets 
to fixed thermal gradients and employing Monte Carlo simulation with 
modified Metropolis algorithm. 
However, Monte-Carlo studies in this direction \cite{mc-muglia-2012} have been limited to spin-flip moves and non-conserved dynamics.
Performing atomistic molecular dynamics (MD) simulations under a temperature gradient 
is tricky due to hydrodynamics preservation criteria.
The heat exchange algorithm works well \cite{frenkel2015}. 
Non-equilibrium MD simulations were performed to calculate Soret coefficient in Lennard-Jones binary mixtures \cite{zimmermann2022}, for investigating phase stability under temperature gradients \cite{liang-prl2013}.
Studies addressing domain coarsening in phase separating systems using MD with a temperature gradient are scarce.
A master-equation approach for wetting phenomena in a thin film with a 
stationary linear temperature profile revealed phase separation in spatial regions with $T>T_c$ ($T_c$ being the critical temperature of phase transition), 
in contrast to bulk behavior \cite{Jaiswal_2013}.
However, most works on phase separation with inhomogeneous temperature fields 
are for either stationary temperature profiles or spatially fixed heat sources \cite{lee2002, Gonnella_2008}. 
Studies on phase separation due to a \textit{moving} heating/cooling source are exceedingly scarce. 
\begin{figure*}
\centering
\makebox[\textwidth]{\includegraphics[width=0.8\paperwidth]{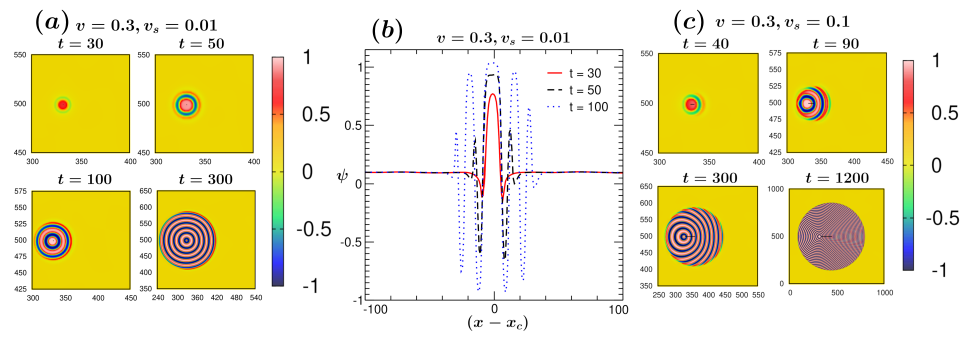}}
\caption{A moving cooling source induced phase separation in a binary mixture 
with an off-critical composition, $\psi_0=0.1$. 
At $t=0$, the source is positioned at $(L/3,L/2)$ and translates along positive x direction with a constant speed $v_s$. 
The cold temperature front propagates radially outward from the source with a constant speed $v$. 
$L=1000$, $dt=0.001$, $T_i=1$, and $T_s=-1$. 
(a): Coarsening patterns for $v=0.3$ and $v_s=0.01$ are shown at four representative times. 
The color code represents the order parameter $\psi$, where $\psi>0$ and $<0$ correspond to the A and B rich phase, respectively. 
Near the source, concentric circular rings of alternate phases emerge.
As time progresses, additional rings form and the overall domain size increases.
Far from the source, the system remains in the mixed state ($\psi=0$). 
For clarity, only a portion of the simulation box is displayed.
(b) The order parameter profile $\psi(x)$ is plotted as a function of distance x from the initial source location at three
distinct times for $v=0.3$ and $v_s=0.01$.
(c) Coarsening patterns for $v=0.3$ and $v_s=0.1$ are shown at four representative times. 
At early times, concentric rings of alternating phases are observed. 
At later times, as the source moves rightward, semi-circular rings form.}
\label{fig1}
\end{figure*}

In this paper, we present numerical investigations of phase separation driven by 
a `moving' cooling source, from which cold temperature fronts propagate radially outward into the surrounding binary mixture. 
Motion of the source introduces two distinct velocity scales into the system: 
one associated with the translational motion of the source, 
and the other corresponding to the propagation speed of the outward-moving thermal front. 
The interplay between these two velocities is found to play a crucial role in determining the resultant pattern morphology. 
\textit{Inter alia}, this competition makes the non-equilibrium dynamics of the system rich and complex. 

When a system initially in a homogeneously mixed state is driven out of equilibrium - typically by a rapid temperature quench across the critical point of 
a continuous phase transition \cite{bray2002theory, sutapa2018} -
it slowly evolves towards a new equilibrium configuration. 
The ensuing transient and inherently non-equilibrium process is referred to as 
phase separation dynamics \cite{puri2009kinetics,bray2002theory, onuki2002phase}.
During this mechanism, domains of similar species emerge and grow over time.
Their morphology and kinetics are strongly influenced by several factors, 
including temperature, concentration, impurities, boundary conditions, 
confinement by surfaces, etc. 
The mechanism of domain growth in `isothermal' systems has been extensively 
explored \cite{sutapa2019-aging, Das_2012, oprisan2017, Gonnella_2020, gerhard2024, processes2025} through analytical theory, experiments, numerical and simulation methods. 
For completeness, we briefly summarize the key findings from these studies. 

The domains / patterns formed exhibit self-similarity; domain patterns at two 
different times follow a scaling despite the increase in average domain size with time. 
The scaling properties are captured as \cite{onuki2002phase}
\begin{equation}
C(r,t) \equiv \tilde C(r/\ell(t)).
\end{equation}
Here, $C(r,t)$ is the two-point equal-time correlation function and $\ell(t)$ is 
the growing lengthscale in the system, viz., the average domain size. 
$r=|r_i-r_j|$ is the distance between two points.
The above mentioned correlation function is calculated as 
\begin{equation} 
\label{eqcor}
C(r,t)=\langle \psi(\vec r_i,t) \psi(\vec r_j,t) \rangle - 
\langle \psi(\vec r_i,t) \rangle \langle \psi(\vec r_j,t) \rangle,
\end{equation}
where, $\psi$ is the relevant order parameter. 
This kind of scaling is associated with fractal like behaviors \cite{bray2002theory, sahu_2025}. 
The average domain size, $\ell(t)$ grows in a power-law manner with a 
growth exponent $\alpha$ as \cite{puri2009kinetics, bray2002theory}
\begin{equation}
\ell(t) \sim t^{\alpha}. 
\end{equation}
The value of $\alpha$ depends on hydrodynamic effects, order parameter conservation, 
space dimensionality, etc. 
For solids, diffusion is the only relevant growth mechanism and yields $\alpha=1/3$ \cite{lifshitz1961kinetics, binder1991spinodal}. 
To the contrary, for fluids various growth regimes are observed \cite{laradji1996molecular, roy2013effects}. 
Following an early-time diffusive regime, a viscous hydrodynamic regime with 
$\alpha =1$ is observed, followed by an inertial hydrodynamic regime with $\alpha=2/3$ \cite{varsha2023}. 
The morphology depends on the quench concentration. 
For a quench at the critical concentration interconnected percolating structures are observed. 
On the other hand, an off-critical quench leads to the nucleation of droplets and 
they grow in size with time \cite{roy2013dynamics, roy2012nucleation}. 
For the latter, even in the presence of hydrodynamics, one observes $\alpha=1/3$ \cite{binder1977theory, tanaka1997new}. 
By changing the density or concentration of the fluid one can thus tune the morphology. 
\begin{figure*}[htbp]
\centering
\makebox[\textwidth]{\includegraphics[width=0.8\paperwidth]{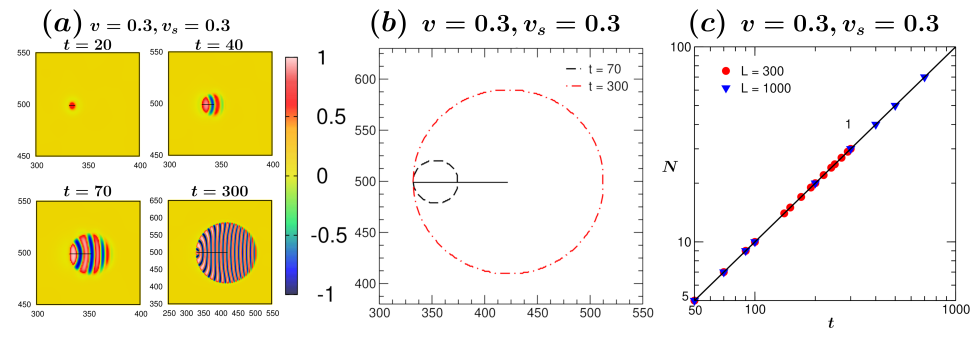}}
\caption{(a) Coarsening patterns for $v=0.3$, $v_s=0.3$ at four representative times are shown for 
$L=1000$, $T_i=1$, $T_s=-1$. 
The position of the moving cooling source is indicated by a solid line.
Circular domains emerge, consisting of stripes of alternating phases.
As time progresses, the domain size increases along with the number of stripes.
(b) Temperature contours corresponding to the same parameters as in (a) are shown at two representative times. 
These contours delineate regions where $a(\vec r,t)>0$ and $<0$.  
(c) Double logarithmic plot of the number of stripes $N$ vs. time $t$ for $v=0.3$, $v_s=0.3$, $T_i=1$, $T_s=-1$ for four different system sizes.
Symbols refer to numerical data and the solid line indicates linear growth.
}
\label{fig2}
\end{figure*}

Presence of temperature gradients significantly alter the pattern morphology and 
their growth dynamics. 
Studies on domain growth under the influence of inhomogeneous temperature fields are  few. 
Phase separation by directional quenching turned out to be a promising method for creating regular material structures \cite{krekhov2009, kurita2017, roy2018colloid}. Dependence of the domain morphology on instabilities and geometry were also studied in \cite{furukawa-physica1992}. Phase separation structures due to enslaved fronts were studied numerically \cite{foard-pre2009}. However, these investigations correspond to spatially fixed heat sources. 
Lattice-Boltzmann (LB) simulations were performed to study phase separation in the presence of cold walls \cite{gonnella-pre2010} and investigate the dependence on viscosity, composition, and diffisivity. LB simulations of the wake of phase separation fronts in parallel geometry demonstrated the presence lamellar patterns \cite{LBpre2012}.
Effects of the interplay between a traveling temperature wave and thermal diffusion was studied by considering a cosine traveling wave function \cite{wave2009}. 
This latter work demonstrated that a strong amplitude of oscillation yields three characteristic velocity regimes in which distinct evolving patterns can be found. 
Stripe patterns were observed by a rotating quench front \cite{prl2006biro}. 
But the influence of spatially mobile heat source with radial heat propagation on the pattern morphology and kinetics remain largely unexplored.  

Rest of the paper is organized as follows: 
In \cref{model}, we explain the model and computational methodologies used. 
\cref{results} present our numerical findings from a systematic study by varying several parameters, viz., average order parameter, source temperature $T_s$, velocity of front propagation $v$, speed of cooling source $v_s$, etc. 
Finally, \cref{conclusions} concludes the paper with a brief summary of the work and presents an outlook and possible future works. 

\section{Model and Methods}\label{model}
We consider a two-dimensional binary mixture in which cold temperature fronts propagate outward from a cooling source that moves through the medium at a constant velocity. 
Propagation of these cold waves yields a time-dependent temperature gradient 
across the system. 
The system has an upper critical temperature $T_c$ \cite{crctemp} of a second-order demixing phase transition \cite{demixing}. Above $T_c$, both phases of the binary mixture are homogeneously mixed and below $T_c$ it phase separates into A- and B- rich phases.
Initially, the entire system is maintained at a fixed temperature higher than $T_c$, ensuring an isothermal system and homogeneously mixed state. 
At time $t=0$, the cooling source with $T_s=-1$ ($<T_c$) is introduced. 
The cold fronts propagate radially outward from the source at a constant speed $v$.  
Simultaneously, the cooling source translates in space with a constant velocity $v_s$. 

\begin{figure}[htbp]
\centering
\includegraphics*[width=0.3\textwidth]{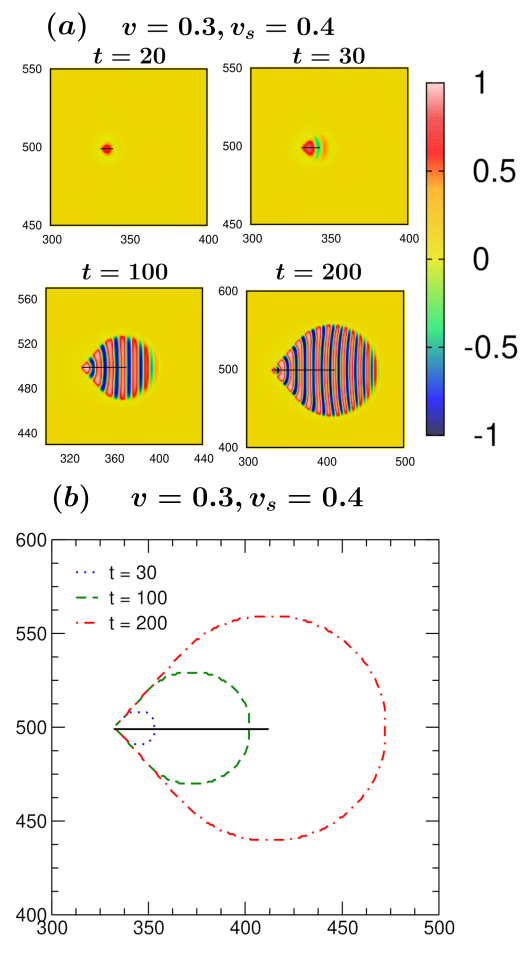}
\caption{
(a) Phase separation patterns for $v=0.3$ and $v_s=0.4$ at four different times are shown.
In the vicinity of the cooling source, an asymmetric, leaf-like domain emerges, consisting of stripes of alternating phases. 
With increasing time, additional stripes form and the overall domain size increases. 
For clarity, only a portion of the simulation domain is displayed.
(b) Temperature contours corresponding to the same parameters as in (a) are shown at three representative times. 
The leaf-shaped contour, arising from the condition $v_s>v$, closely correlates with the domain morphology. 
As time progresses, the source moves rightward and the contour becomes larger.
}
\label{fig3}
\end{figure}

Competition between $v_s$ and $v$ dictates temperature field across the system which directly controls the spatiotemporal pattern formation. 
When $v_s<<v$, the cooling front spreads over an extended region before the source moves.
In contrast, for comparable velocities $v_s \sim v$, the temperature gradients generated by the moving source drive directional phase separation and lead to anisotropic patterns. 
When the source outruns thermal diffusion $v_s>>v$, the source moves forward before the thermal front has obtained sufficient time to expand in space.

In order to investigate the time evolution of the temperature field and the resulting phase separation patterns, we utilize the phenomenological framework of Cahn-Hilliard-Cook (CHC) model \cite{CHC}, originally derived from the Landau-Ginzburg free energy functional \cite{func}:
\begin{equation}
\label{free}
\frac{F}{k_BT_c} = \frac{1}{\nu_0} \int d^{3}r \Bigl[ \frac{1}{2} a{\psi}(r)^{2}
+ \frac{1}{4} b{\psi}(r)^{4} + \frac{1}{2} C\bigl({\nabla} {\psi}(r)\bigr)^{2}\Bigr],
\end{equation} 
here, $F$ and $\psi$ are the free energy and order parameter, respectively. For a binary mixture, $\psi$ is the difference in concentration of A and B species. Positive and negative values of $\psi$ would correspond to the A and B phases, respectively. Also, $\nu_0$ is a volume unit and $a<0$, $b>0$, $C>0$ are constants. 

Consideration of order parameter conservation and concentration current $\boldsymbol j$ yields \cite{func}
\begin{equation}
 \label{eq:chc1}
 \frac{\partial {\psi}(r,{t})}{\partial {t}} = 
 -{\nabla} \cdot {\boldsymbol j}({r},{t})= -\nabla \cdot (-M \nabla \mu)=M\nabla^2 \mu= M\nabla^2 \Bigl[ \frac{\partial F}{\partial \psi}\Bigr].
\end{equation}
Here, $M$ and $\mu =\partial F / \partial \psi$ are the mobility and chemical potential, respectively \cite{CHC}.
With appropriate rescaling factors
\begin{equation}
\label{rscal}
    \tilde{\psi} = \sqrt{\frac{a}{b}},\quad
       \tilde{r} = \sqrt{\frac{C}{a}},\quad
    \tilde{t} = \frac{\nu_0\tilde{r}^2}{Mk_BT_ca}
\end{equation}
\cref{free,eq:chc1,rscal} yields the non-dimensionalized Cahn-Hilliard-Cook (CHC) equation \cite{CHC}

\begin{equation}\label{scaledchc}
\frac{\partial \psi(\vec r,t)}{\partial t} = M\nabla ^2 \Big ( a\psi(\vec r,t) + \psi^3(\vec r,t) - \nabla^2 \psi(\vec r,t) \Big) +\eta(\vec r,t).
\end{equation}
Here, we have considered a Gaussian random noise $\eta(\vec r,t)$ \cite{CHC} which obeys the relation $\langle \eta(\vec r,t)\eta(\vec r',t') \rangle= -2\nu(r)\nabla^2\delta(\vec r - \vec r')\delta(t - t') $, $\nu(\vec r)$ characterizing the noise strength. In our work, there is a temperature field which depends on both space and time. In order to incorporate this, we make the parameter $a$ in \cref{scaledchc} space dependent: $a=a_0 (T(\vec r,t)-T_c)/T_c$, $a_0$ being a positive constant. By construction, $T(\vec r,t)<T_c$ favors phase separation and yields $a(\vec r,t)<0$.

Initially, we set $a(\vec r,t)=1$ everywhere in the system. 
For the spatial region inside which the cold front has propagated, 
$|\vec r-\vec r_c|\le vt$, $a(\vec r,t)<0$ and it favors phase separation. 
Here, $\vec r_c$ is the location of the cooling source at time $t=0$. 
Whereas, regions with $|\vec r-\vec r_c|>vt$ are associated with positive $a(\vec r,t)$ and prefers a homogeneously mixed state. 
This gives rise to a local phase separation. 
It would be worthwhile investigating whether phase separation stays restricted only into the region with $a(\vec r, t)<0$ or extends beyond it. 
Simultaneously, the cooling source moves through space with a constant velocity $v_s$. This gives rise to a dynamic coupling between the two characteristic velocities, local concentration, and local temperature field. 
The resulting temperature distribution is therefore both time-dependent and spatially inhomogeneous. 
The patterns generated turn out to be strongly governed by the velocity ratio $v_s/v=\gamma$. 
Apriori it is difficult to predict the influence of $\gamma$ on the pattern morphology. 

We performed numerical calculations using square box of side length $L$ (two-dimensions), subjected to fixed boundary conditions (b.c.) imposed at the outer edges of the computational box. 
The temperature at the outer boundaries is maintained at $T_{\text{boundary}}=T_i$. 
This ensures that the influence of the moving cooling source remains limited only to the interior of the simulation box. 
This corresponds to a large sample cell in which the effects of propagating cold fronts almost do not reach the edge of the cell.
A finite-difference method is used to solve \cref{eq:chc1}.  
The time step of integration $dt=0.001$. $M$ is taken to be constant (unity) and $\nu=0.00005$.

\section{Results}\label{results}
\subsection{Off-critical composition}

We begin by presenting results for an off-critical average order parameter $\psi_0 =0.1$. 
$\psi_0$ is defined as $(1/N)\sum_{i=1}^N \psi_i$, where $i$ stands for a grid point of the square simulation box and $N$ is its total number.
For this velocity, we analyze the spatiotemporal pattern evolution across different parameter sets.
\cref{fig1}(a) shows the coarsening patterns at four distinct times for the velocity combination $v=0.3$ and $v_s=0.01$. 
System parameters correspond to $L=1000$, $T_i=1$, $T_s=-1$, $dt=0.001$. 
At time $t=0$, the cooling source is kept at location $(L/3,L/2)$. 
It translates along the positive $x$- direction. 
Color bar corresponds to the full range of order parameter $\psi$, where $\psi>0$ and $\psi<0$ stand for phase A and B, respectively.
For better clarity, only a portion of the simulation box is displayed. 
At all times, patterns consist of concentric circular rings. 
Two consecutive rings are formed of opposite phases. 
At $t=30$, a single layer characterized by $\psi>0$ has formed, which is encompassed by an adjacent layer of the alternate phase $\psi<0$. 
With increasing time ($t=100$ and $300$), additional rings develop. 
Patterns develop along the direction of propagation of the cooling source.
Away from the source, the system stays in the mixed state with $\psi=0$ (yellow color).
Phase separation is thus `local' in nature. Note that at this very small $v_s$ the cooling source remains nearly stationary while the temperature front propagates radially outward, resulting in circularly symmetric patterns. 
It is anticipated that larger values of $v_s$ will disrupt this symmetry.

In \cref{fig1}(b), the order parameter profile $\psi(x)$ is presented as a function of the distance $(x-x_c)$ from the initial source location $x_c$, at three representative times.  
System parameters are same as \cref{fig1}(a). 
At $t=30$, a layer of phase A ($\psi>0$) forms adjacent to the source at $x=x_c$, followed by a neighboring layer of phase B ($\psi>0$). For  $x>x_c+20$, no significant structure is observed. As time increases, additional layers emerge.

Next, we perform a systematic investigation of the dependence of the pattern morphology and the kinetics on $v_s$, while keeping $v$ fixed. 
For this, in \cref{fig1}(c) we present evolution snapshots for $v=0.3$, $v_s=0.1$ at four distinct times. 
At early times, circular rings of alternate phases are observed. 
At later times, semicircular stripes develop, reflecting the rightward motion of the cooling source. These stripes exhibit significant curvature and a concavo–convex geometry. The increase in $v_s$ thus induces a transition from symmetric patterns to asymmetric ones.
\begin{figure}[t]
\centering
\includegraphics*[width=0.4\textwidth]{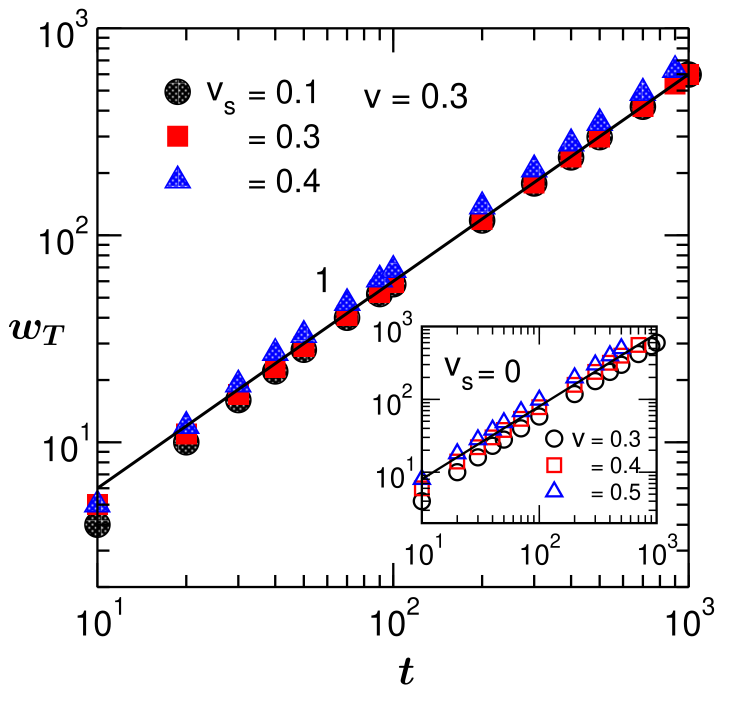}
\caption{
(a) Plot of the spatial region $w_T$ over which the cold temperature front propagates, defined by the condition $a(\vec r,t)<0$, vs. time $t$, on a double-logarithmic scale. 
Solid line stands for a linear growth. 
Inset: same plot for a stationary source ($v_s=0$). 
}
\label{fig4}
\end{figure}

Next, we further increase the source velocity to $v_s=0.3$ ($v=0.3$). 
The corresponding evolution snapshots are shown in \cref{fig2}(a). 
For clarity, only a portion of the simulation box is displayed and the position of the cooling source is marked with a black solid line.  
The resulting patterns no longer hold concentric rings; instead, vertical stripe-like structures emerge within a circular domain. 
As time progresses, both the size of the domain and the number of stripes increase. 

In \cref{fig2}(b), we present the temperature contour across the simulation box, delineating regions where $a(\vec r,t)>0$ and $<0$. 
Systems parameters are same as \cref{fig2}(a). 
Notably, the contour has a circular shape. 
This is because the cooling source moves through the system at the same rate at which the temperature front spreads outward. 
For a larger $v_s$, the source would outrun the front —leading to asymmetry in the temperature contours.

To further characterize the patterns, in \cref{fig2}(c) we plot the number $N$ of stripes vs. time $t$, on a double-logarithmic scale for different system sizes $L$. 
The symbols denote numerical results, which are in good agreement with the linear growth ($\beta=1$) indicated by the solid line.

In \cref{fig3}, we present results for $v=0.3$, $v_s=0.4$. 
The pattern evolution is depicted in \cref{fig3}(a). 
The domains exhibit an asymmetric, \textit{leaf-like} morphology, within which vertical stripe-like structures emerge. 
As time progresses, the cooling source translates rightward (indicated by the solid black line), and the domain size increases. Successive stripes always consist of alternating phases.
The inner stripes display a concavo–convex geometry, whereas the outer stripes, at later times, approach a plano–convex form. 
Far from the source, the system remains in a mixed state with $\psi=0$ (yellow color).

\cref{fig3}(b) shows the temperature contour across the simulation box at three different times, delineating regions with $a(\vec r,t)>0$ and $<0$, for $v_s=0.4, v=0.3$. 
The contour has an asymmetric shape. 
Since the source moves faster than the temperature front can diffuse ($v_s > v$), 
the cold region extends rightward more rapidly than it can spread radially from its earlier positions. 
This results in an asymmetric temperature profile, which in turn gives rise to an asymmetric phase separating domain. 
With increasing time, the contour gets bigger.

For $v=0.3, v_s=0.3$ (Fig. 2), both the contour and the phase separation pattern are circular. 
For $v=0.3, v_s=0.4$ (Fig. 3), both exhibit leaf-like shape. 
This suggests that asymmetry in the temperature contour directly influences the asymmetry of the coarsening morphology. 
The effect arises from the localized nature of phase separation, where the contour reflects the interplay between
$v_s$ and $v$.

\begin{figure}[htbp]
\centering
\includegraphics*[width=0.4\textwidth]{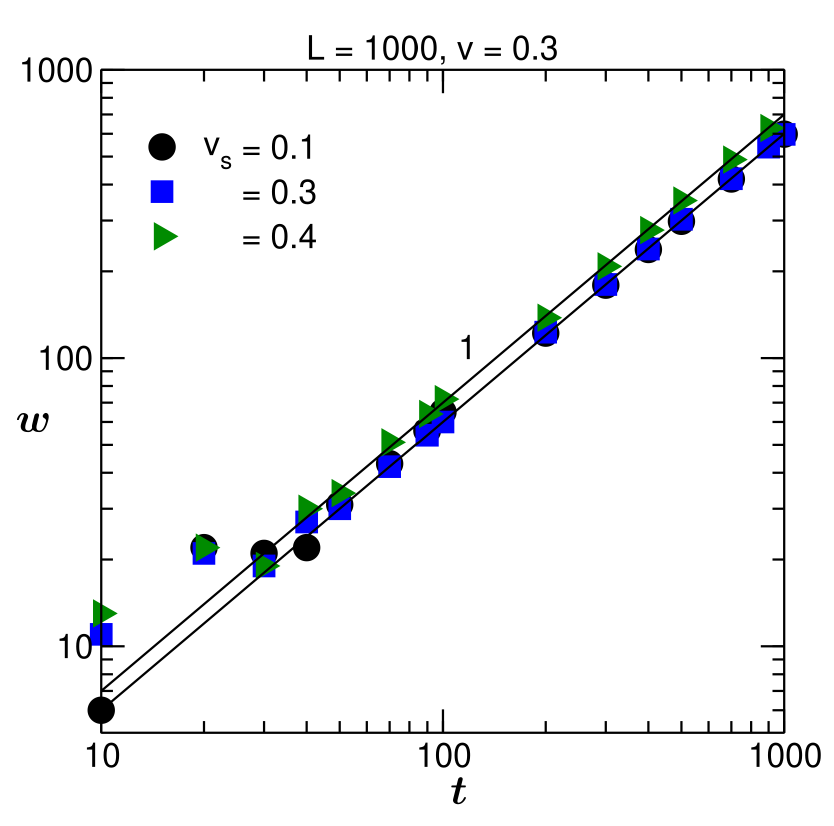}
\caption{
Log–log plot of the domain width $w$ as a function of time for fixed $v$ and three different $v_s$. 
}
\label{fig5}
\end{figure}

To elucidate the growth kinetics, \cref{fig4} illustrates the temporal evolution of the spatial region $w_T$ over which the cold temperature front propagates, defined by the condition $a(\vec r,t)<0$. For a fixed front velocity $v$ and three distinct values of $v_s$, $w_T$ exhibits a clear linear dependence on time $t$. A similar linear growth behavior is observed even when the source remains stationary (inset).

In \cref{fig5}, we demonstrate the growth of the domain width $w$ which is measured as the longitudinal extent of the domain along the direction of motion of the cooling source (i.e., x axis at $y=L/2$). Data for $w$ corroborate well with a linear growth marked by solid lines. When the cooling source moves along the x-axis, the resulting patterns remain symmetric with respect to $y=L/2$ and the extension $w$ is effectively uni-dimensional. If the source were to move randomly in the x-y plane, $w$ would no longer exhibit a uni-directional character.  

\begin{figure}[htbp]
\centering
\includegraphics*[width=0.5\textwidth]{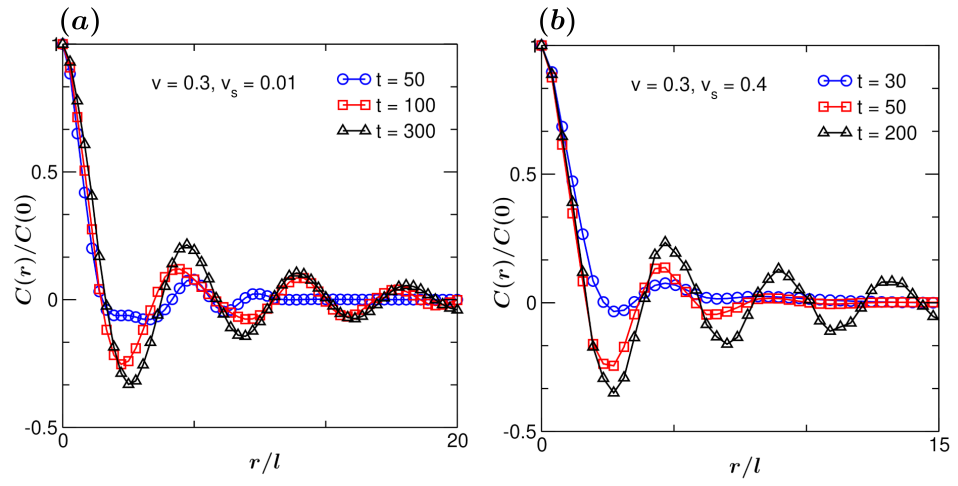}
\caption{Scaled two-point equal time correlation function $C(r,t)/C(r=0,t)$, is shown at three representative times for $v=0.3$ and two different values of $v_s$. Data from different times are marked by different symbols. 
}
\label{fig6}
\end{figure}
In isothermal systems, phase-separation patterns typically exhibit self-similarity, reflected in the scaling of $C(r,t)$ (see \cref{eqcor}). In \cref{fig6}, we show the scaled two-point correlation function $C(r,t)$ vs. $r/\ell$ ($\ell$ being the domain size) for two distinct cases $v=0.3, v_s=0.01$ ($v_s<v$) and $v=0.3, v_s=0.4$ ($v_s>v$). The length scale $l(t)$ is defined as the distance $r$ at which the correlation function $C(r,t)$ decays to $0.5$ of its maximum value. Interestingly, in both cases, data collapse was not observed. This indicates the non self-similar nature of the domain patterns. 

\begin{figure}
\centering
\includegraphics*[width=0.5\textwidth]{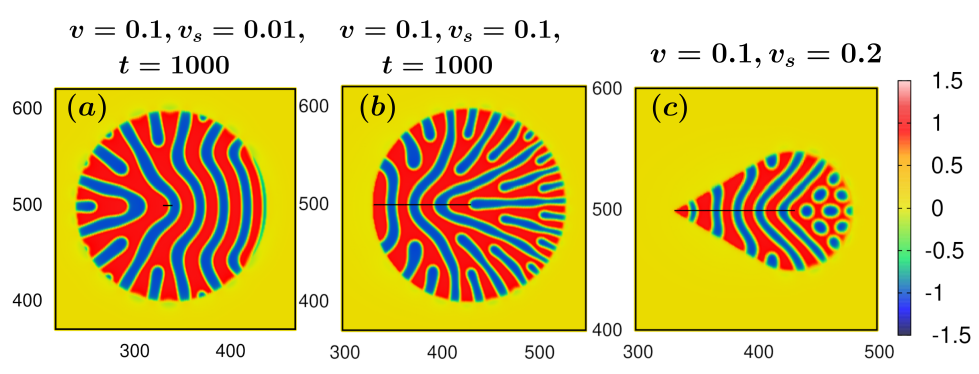}
\caption{ 
Domain morphology for (a) $v=0.1$, $v_s=0.01$ and (b) $v=0.1$, $v_s=0.1$, for $L=1000$, $T_i=1$, $T_s=-1$.
Translational speed of the cooling source $v_s$ influences the direction of bending of the stripes.
(c) Domain morphology for $v=0.1$ and $v_s=0.2$ is shown at a representative time. 
The significantly larger source velocity $v_s$ (as compared to the  front velocity $v$) yields strongly asymmetric pattern.
}
\label{fig7}
\end{figure}

In \cref{fig7}(a), we explore the dependence of morphology on the source velocity $v_s$, for a fixed $v$. 
Representative snaphots are shown here. 
In both cases, the domains are circular and consist of stripes. 
However, the curvature of the stripes is opposite in the two cases: they bend toward the left for $v_s=0.01$ and toward the right for $v_s=0.1$.
The velocity ratio $v_s/v$ thus controls the pattern morphology, including its asymmetry, shape, and size. 

We also examined the effect of changing $v$ on the coarsening patterns, for a fixed $v_s=0.2$. 
For $v=0.1, v_s=0.2$, strongly asymmetric pattern forms (presented in \cref{fig7}(c)).
This asymmetry arises because $v_s>v$.
For $v=0.2, v_s=0.2$ circular domain with vertical stripes are observed (similar to \cref{fig2}(a)). 
For $v=0.6, v_s=0.2$ circular domain with semicircular strips form (similar to \cref{fig1}(c)). 
Upon further increasing $v$, for $v=0.8,v_s=0.2$ $(v_s << v)$, concentric circular rings are observed (similar to \cref{fig1}(a).) 

Our results reveal that the coarsening morphology depends not only on the velocity ratio $\gamma=v_s/v$, but also on the absolute magnitudes of $v_s$ and $v$. 
The domain morphology for $v_s=0.1, v=0.1$ ($\gamma=1$, \cref{fig7}) contains lamellar stripes. 
To the contrary, for $v_s=0.3, v=0.3$ ($\gamma=1$, \cref{fig2}) vertical stripes are observed.

\begin{figure}
\centering
\includegraphics*[width=0.4\textwidth]{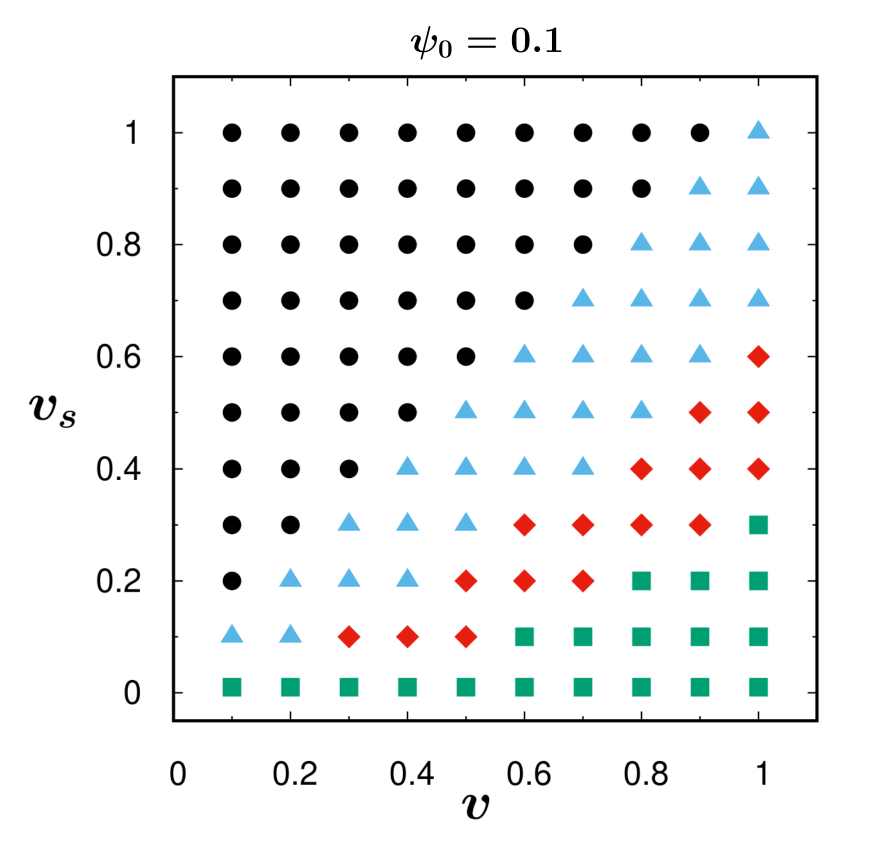}
\caption{  
Phase diagram of the system in the $v-v_s$ parameter plane for $\psi_0 = 0.1$. 
All data refer to $T_i=1$, $T_s=-1$. 
Symbols correspond to distinct pattern shapes- square: concentric circular rings, diamond: semicircular stripes pattern, triangle up: circular domain with vertical stripes, circle: asymmetric domain with stripes. }
\label{fig8}
\end{figure}

In \cref{fig8}, we present the phase diagram of the system in the $v-v_s$ parameter space. 
Distinct morphologies are marked with different symbols. 
A comprehensive understanding of this parameter space will enable an efficient engineering of pattern morphology on substrates through controlled motion of the cooling/heating source, i.e., by tuning the two velocities $v$ and $v_s$. 

\begin{figure}
\centering
\includegraphics*[width=0.4\textwidth]{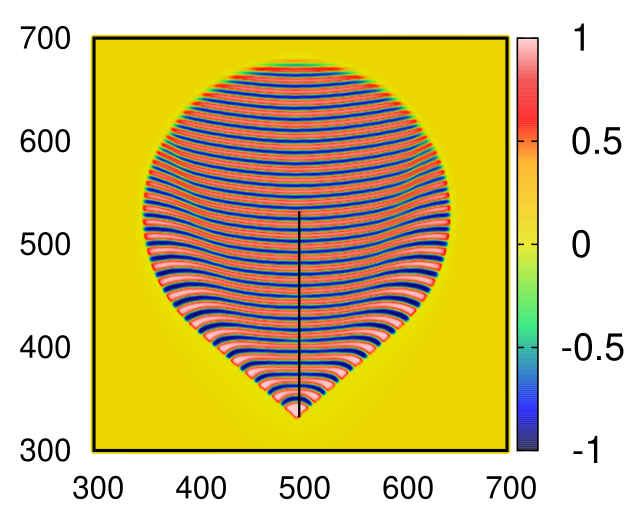}
\caption{Domain morphology for $v=0.3$, $v_s=0.4$ for a cooling source moving along the y direction. Initially the source is placed at $(L/2,L/3)$. 
The solid line represents the propagation of the cooling source.
}
\label{fig9}
\end{figure}

To assess the influence of the source direction on phase-separation patterns, we examine the case where the source moves along the y-direction. 
In \cref{fig9}, we present a representative snapshot for $v=0.3$ and $v_s=0.4$. 
The source is initially positioned at ($L/2,L/3$). 
A comparison with the case of motion along the x-direction reveals that the qualitative features of the domains remain largely unchanged, indicating that the pattern type is robust to the direction of source motion. However, direction of motion influences the orientation of the stripes.

\subsection{Critical composition:}
\begin{figure}
\centering
\includegraphics*[width=0.5\textwidth]{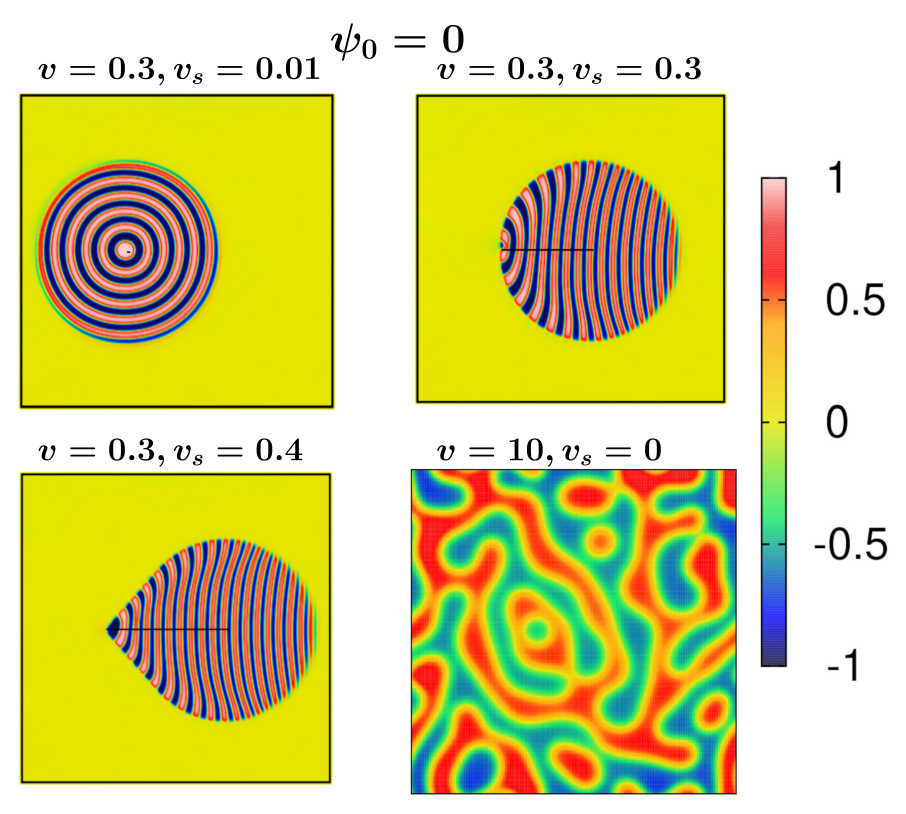}
\caption{ Pattern morphologies at critical composition $\psi_0=0$, for different combinations of $v$ and $v_s$.
}
\label{fig10}
\end{figure}

For the critical composition $\psi_0=0$, the domain morphologies are observed to be qualitatively similar to those for $\psi_0=0.1$, including concentric circular rings, semi-circular rings, circular domain with vertical stripes, asymmetric domain with stripes are found. Representative snapshots for various combinations of $v$ and $v_s$ are shown in \cref{fig10} (a-c). Notably, at very large $v=10$ with $v_s=0$, a bicontinuous structure characteristic of spinodal decomposition emerges, which is absent for $\psi_0=0.1$.

\begin{figure}
\centering
\includegraphics*[width=0.5\textwidth]{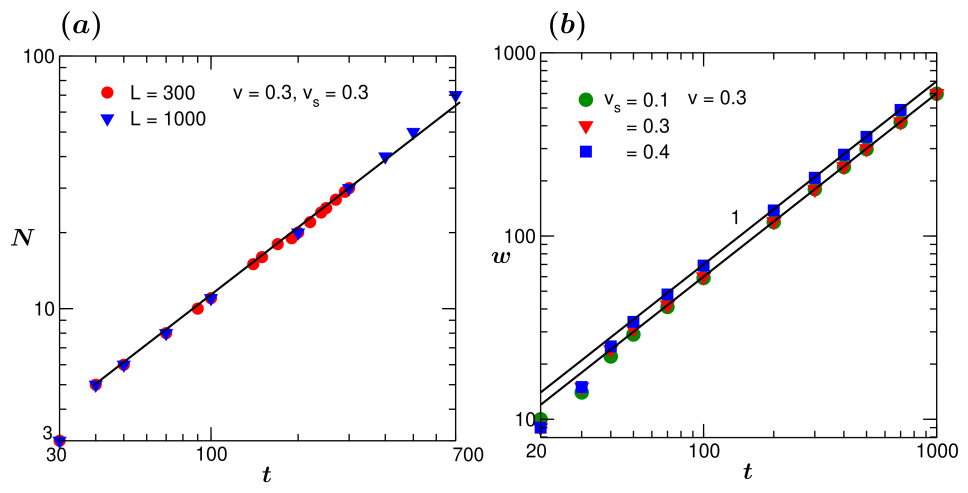}
\caption{ (a) Time-dependence of the number of stripes within a domain for $v=0.3,v_s=0.3$, on a log-log plot. (b) Growth of the width $w$ of the domain as a function of time $t$, for a fixed $v=0.3$ and three different values of $v_s$. Solid lines mark linear growth. For both (a) and (b), $\psi_0=0$.  
}
\label{fig11}
\end{figure}

In \cref{fig11}(a), we demonstrate the growth of the number of stripes within a domain vs. time, for $v=0.3, v_s=0.3$. Numerical data for two different system sizes corroborate well with a power-law growth behavior marked by the solid line. 

Time dependence of the domain width $w$ is shown in \cref{fig11}(b), on a double-logarithmic scale from $L=1000$ for fixed $v=0.3$ and three different values of $v_s$. $w$ which is measured as the longitudinal extent of the domain along the direction of motion of the cooling source (i.e., x axis at $y=L/2$). For all $v_s$, our numerical data exhibit a linear growth marked by the solid lines. 

In the present work, we have not accounted for the effects of surfaces. 
In practice, surfaces often exhibit preferential affinity toward one component of a binary mixture, leading to surface adsorption and surface-directed spinodal decomposition (SDSD) \cite{das2006molecular}. It will be interesting to investigate the effects of thermal gradient for phase separation in confined geometry, interface structure \cite{felix2024-jcp}, two-phase flow in nanoporous media \cite{fayaz2022}. 

\section{Conclusions}\label{conclusions}
In summary, we have studied the non-equilibrium phase separation in a binary mixture driven by a moving cooling source. 
The cooling source has a temperature $T_s$ below the upper critical temperature $T_c$ of a demixing phase transition. 
Motion of the cooling source introduces \textit{two} distinct velocity scales in the system: translational motion of the source $v_s$, and propagation $v$ of outward-moving cooling fronts.
There is a competition between the motion of the cooling source $v_s$ and the propagation of cold wave in the fluid which is characterized by an effective front velocity $v$. 
This competition determines how long a region of the fluid experiences a temperature change, which directly controls whether phase separation can occur or not and how domains develop. 

In our work, a Cahn-Hilliard-Cook (CHC) model has been used which appropriately couples the time-dependent temperature field and concentration field. 
We numerically solve the resultant CHC equation using the finite-difference method. 
Our study reveals that the coarsening morphology and its kinetics strongly depend upon the ratio of two velocities $\gamma=v_s/v$ and their absolute values. 
Same value of $\gamma$ yields distinctly different morphologies for $v=0.3$ and $0.1$. 
Source speed can alter the direction of bending of the coarsening stripes. 
Local phase separation is observed. 
As opposed to isothermal fluids, self-similarity is not observed in these patterns. 
The temperature contour delineating the region with local temperature larger than the phase transition temperature $T_c$ and $<T_c$ controls the shape of the coarsening domain. 

Here we summarize our observations for different velocity ranges.
\textbf{Case I}: Slow source $v_s << v$: cold temperature front spreads significantly faster than the source, preventing the source from imprinting any directional signature on the pattern. 
Temperature field quickly equilibrates and the system experiences standard spinodal decomposition under a nearly uniform quench.

\textbf{Case II}: Comparable velocities $v_s \sim v$: spreading of the cold temperature front and the source motion occur on similar time scales, leading to an asymmetric temperature field. 
As a result, the system is constantly driven out of equilibrium and the domains preferentially align along the direction of source motion. 
This regime produces the richest pattern morphology.

\textbf{Case III}: Fast source $v_s >> v$: source moves much faster than temperature front can diffuse. In this case, the medium does not get sufficient time to develop for large phase separated structures. 
Consequently, the domains remain small and of elongated shapes.

Studies on phase separation under inhomogeneous temperature fields are less. 
The existing literature is mostly dedicated to spatially fixed heating/cooling sources and with stationary temperature profiles.
Our study contributes to the understanding of coarsening induced by a `moving' cooling source. 
It is also relevant for structure formation around moving \textit{laser} illumination \cite{Gomezprl2016}. 
Our results provide a framework for efficient design of pattern morphologies on substrates through controlled manipulation of the cooling/heating source, specifically by tuning the velocities $v$ and $v_s$. 
We believe this will lead to further experimental works in this direction. 
From an experimental perspective, temperature control in confined liquid systems is typically imposed at the base of the sample cell rather than uniformly throughout the bulk. 
This underscores the need for theoretical models that describe phase separation in systems confined between boundaries maintained at different temperatures. We hope our work will motivate further studies in the latter direction.

\vskip 1.0cm
Acknowledgement: SR acknowledges financial support from BITS Pilani, India, through NFSG grant no: N5/25/1053.

\newpage

\bibliography{phasesep}

@Article{processes2025,
AUTHOR = {Li, Wenbin and Hu, Jie and Liu, Renxin},
TITLE = {Displacement Mechanism of Sequential Droplets on a Wetting Confinement},
JOURNAL = {Processes},
VOLUME = {13},
YEAR = {2025},
NUMBER = {9},
ARTICLE-NUMBER = {3014},
URL = {https://www.mdpi.com/2227-9717/13/9/3014},
ISSN = {2227-9717},
DOI = {10.3390/pr13093014}
}

@article{gerhard2024,
    author = {Biswas, Tanmay and Kahl, Gerhard and Shrivastav, Gaurav P.},
    title = {Phase separation dynamics in a symmetric binary mixture of ultrasoft particles},
    journal = {The Journal of Chemical Physics},
    volume = {160},
    number = {21},
    pages = {214901},
    year = {2024},
    month = {06},
    issn = {0021-9606},
    doi = {10.1063/5.0209814},
    url = {https://doi.org/10.1063/5.0209814},
    eprint = {https://pubs.aip.org/aip/jcp/article-pdf/doi/10.1063/5.0209814/19972444/214901_1_5.0209814.pdf},
}

@article{Gonnella_2020,
doi = {10.1088/1751-8121/ab966f},
url = {https://doi.org/10.1088/1751-8121/ab966f},
year = {2020},
month = {jul},
publisher = {IOP Publishing},
volume = {53},
number = {30},
pages = {305002},
author = {Gonnella, G and Lamura, A},
title = {Sheared phase-separating binary mixtures with surface diffusion},
journal = {Journal of Physics A: Mathematical and Theoretical}
}

@Article{oprisan2017,
AUTHOR = {Oprisan, Ana and Garrabos, Yves and Lecoutre, Carole and Beysens, Daniel},
TITLE = {Pattern Evolution during Double Liquid-Vapor Phase Transitions under Weightlessness},
JOURNAL = {Molecules},
VOLUME = {22},
YEAR = {2017},
NUMBER = {6},
ARTICLE-NUMBER = {947},
URL = {https://www.mdpi.com/1420-3049/22/6/947},
PubMedID = {28598367},
ISSN = {1420-3049},
DOI = {10.3390/molecules22060947}
}

@Article{varsha2023,
author ="Singh, Anuj Kumar and Banerjee, Varsha",
title  ="Accelerated inertial regime in the spinodal decomposition of magnetic fluids",
journal  ="Soft Matter",
year  ="2023",
volume  ="19",
issue  ="13",
pages  ="2370-2376",
publisher  ="The Royal Society of Chemistry",
doi  ="10.1039/D3SM00285C",
url  ="http://dx.doi.org/10.1039/D3SM00285C"}

@Article{matsui2023,
author ="Matsui, Yohei and Maeda, Yuki and Kawase, Makoto and Suzuki, Takahiro and Tsushima, Shohji",
title  ="Flow battery recharging by thermoresponsive liquid–liquid phase separation",
journal  ="Sustainable Energy Fuels",
year  ="2023",
volume  ="7",
issue  ="16",
pages  ="3832-3841",
publisher  ="The Royal Society of Chemistry",
doi  ="10.1039/D3SE00451A",
url  ="http://dx.doi.org/10.1039/D3SE00451A"}

@article{zhou2025,
    author = {Zhou, Hao and Jin, Ziguan and Xu, Yuhong and Lu, Yuyao and Xia, Zhuoheng and Yang, Fan and Wu, Qianglong and Gao, Yang and Yin, Jun and Zhang, Jianhua and Ni, Chujun and Zhang, Bin and He, Yong and Yang, Huayong and Xu, Kaichen},
    title = {Enhanced laser-induced PEDOT-based hydrogels for highly conductive bioelectronics},
    journal = {National Science Review},
    volume = {12},
    number = {6},
    pages = {nwaf136},
    year = {2025},
    month = {04},
    issn = {2095-5138},
    doi = {10.1093/nsr/nwaf136},
    url = {https://doi.org/10.1093/nsr/nwaf136},
    eprint = {https://academic.oup.com/nsr/article-pdf/12/6/nwaf136/62872643/nwaf136.pdf}
}

@article{ktafi2024,
title = {Crystalline / glass nanoscale chemical separation induced by femtosecond laser pulses in aluminosilicate glass},
journal = {Optical Materials},
volume = {150},
pages = {115294},
year = {2024},
issn = {0925-3467},
doi = {https://doi.org/10.1016/j.optmat.2024.115294},
url = {https://www.sciencedirect.com/science/article/pii/S0925346724004774},
author = {Imane Ktafi and Matthieu Lancry and Marc Dussauze and Bertrand Poumellec and Yasuhiko Shimotsuma and Daniel R. Neuville and Maxime Vallet and Louis Cornet and Maxime Cavillon}
}

@article{amaya2025,
author = {Jim{\'e}nez
Amaya, Ana and Hill, Eric H.},
title = {Perspective: Thermophoresis and Its Promise for Optical Patterning},
journal = {Langmuir},
volume = {41},
number = {21},
pages = {12835-12840},
year = {2025},
doi = {10.1021/acs.langmuir.5c01023},
URL = { 
    
        https://doi.org/10.1021/acs.langmuir.5c01023
   },
eprint = { 
    
        https://doi.org/10.1021/acs.langmuir.5c01023
    
    

}

}

@Article{sutapa2019-aging,
author ="Roy, Sutapa and Bera, Arabinda and Majumder, Suman and Das, Subir K.",
title  ="Aging phenomena during phase separation in fluids: decay of autocorrelation for vapor–liquid transitions",
journal  ="Soft Matter",
year  ="2019",
volume  ="15",
issue  ="23",
pages  ="4743-4750",
publisher  ="The Royal Society of Chemistry",
doi  ="10.1039/C9SM00366E",
url  ="http://dx.doi.org/10.1039/C9SM00366E"}

@Article{sutapa2018,
author ="Roy, Sutapa and Das, Subir K.",
title  ="Study of critical dynamics in fluids via molecular dynamics in canonical ensemble",
journal  ="The European Physical Journal E",
year  ="2018",
volume  ="38",
issue  ="12",
pages  ="132",
doi  ="10.1140/epje/i2015-15132-2",
url  ="https://doi.org/10.1140/epje/i2015-15132-2",
}

@Article{roy2018colloid,
author ="Roy, Sutapa and Maciołek, Anna",
title  ="Phase separation around a heated colloid in bulk and under confinement",
journal  ="Soft Matter",
year  ="2018",
volume  ="14",
issue  ="46",
pages  ="9326-9335",
publisher  ="The Royal Society of Chemistry",
doi  ="10.1039/C8SM01258J",
url  ="http://dx.doi.org/10.1039/C8SM01258J"}

@article{Das_2012,
doi = {10.1209/0295-5075/97/66006},
url = {https://doi.org/10.1209/0295-5075/97/66006},
year = {2012},
month = {mar},
publisher = {},
volume = {97},
number = {6},
pages = {66006},
author = {Das, Subir K. and Roy, Sutapa and Majumder, Suman and Ahmad, Shaista},
title = {Finite-size effects in dynamics: Critical vs. coarsening phenomena},
journal = {Europhysics Letters}
}

@article{das2006molecular,
  author    = {Das, S. K. and Puri, S. and Horbach, J. and Binder, K.},
  title     = {Molecular Dynamics Study of Phase Separation Kinetics in Thin Films},
  journal   = {Physical Review Letters},
  volume    = {96},
  number    = {1},
  pages     = {016107},
  year      = {2006},
  publisher = {American Physical Society},
  doi       = {10.1103/PhysRevLett.96.016107}
}

@article{roy2013effects,
  author    = {Roy, S. and Das, S. K.},
  title     = {Effects of Domain Morphology on Kinetics of Fluid Phase Separation},
  journal   = {Journal of Chemical Physics},
  volume    = {139},
  number    = {4},
  pages     = {044911},
  year      = {2013},
  publisher = {AIP Publishing},
  doi       = {10.1063/1.4816332}
}

@article{roy2013dynamics,
  author    = {Roy, S. and Das, S. K.},
  title     = {Dynamics and Growth of Droplets Close to the Coexistence Curve in Fluids},
  journal   = {Soft Matter},
  volume    = {9},
  number    = {15},
  pages     = {4178--4187},
  year      = {2013},
  publisher = {Royal Society of Chemistry},
  doi       = {10.1039/C3SM27702E}
}

@article{roy2012nucleation,
  author    = {Roy, S. and Das, S. K.},
  title     = {Nucleation and Growth of Droplets in Vapor--Liquid Transitions},
  journal   = {Physical Review E},
  volume    = {85},
  number    = {5},
  pages     = {050602},
  year      = {2012},
  publisher = {American Physical Society},
  doi       = {10.1103/PhysRevE.85.050602}
}

@article{laradji1996molecular,
  author    = {Laradji, M. and Toxvaerd, S. and Mouritsen, O. G.},
  title     = {Molecular Dynamics Simulation of Spinodal Decomposition in Three-Dimensional Binary Fluids},
  journal   = {Physical Review Letters},
  volume    = {77},
  number    = {11},
  pages     = {2253--2256},
  year      = {1996},
  publisher = {American Physical Society},
  doi       = {10.1103/PhysRevLett.77.2253}
}

@article{tanaka1997new,
  author    = {Tanaka, H.},
  title     = {New Mechanisms of Droplet Coarsening in Phase-Separating Fluid Mixtures},
  journal   = {Journal of Chemical Physics},
  volume    = {107},
  number    = {9},
  pages     = {3734--3737},
  year      = {1997},
  doi       = {10.1063/1.474734}
}

@article{binder1977theory,
  author    = {Binder, K.},
  title     = {Theory for the Dynamics of ``Clusters.'' II. Critical Diffusion in Binary Systems and the Kinetics of Phase Separation},
  journal   = {Physical Review B},
  volume    = {15},
  number    = {8},
  pages     = {4425--4447},
  year      = {1977},
  publisher = {American Physical Society},
  doi       = {10.1103/PhysRevB.15.4425}
}

@article{lifshitz1961kinetics,
  author    = {Lifshitz, I. M. and Slyozov, V. V.},
  title     = {The Kinetics of Precipitation from Supersaturated Solid Solutions},
  journal   = {Journal of Physics and Chemistry of Solids},
  volume    = {19},
  pages     = {35--50},
  year      = {1961},
  doi       = {10.1016/0022-3697(61)90054-3}
}

@article{condensate2025,
  author    = {Jin, K. and Yu, W. and Liu, Y. and Li, J. and Du, G. and Chen, J. and Liu, L. and Lv, X.}, 
  title     = {Light-induced programmable solid-liquid phase transition of biomolecular condensates for improved biosynthesis.},
  journal   = {Trends in biotechnology},
  volume    = {43},
  pages     = {1403–1424},
  year      = {2025},
  doi       = {https://doi.org/10.1016/j.tibtech.2025.02.012}
}

@book{puri2009kinetics,
  editor    = {Puri, S. and Wadhawan, V.},
  title     = {Kinetics of Phase Transitions},
  publisher = {CRC Press},
  address   = {Boca Raton},
  year      = {2009}
}

@article{bray2002theory,
  author    = {Bray, A. J.},
  title     = {Theory of Phase-Ordering Kinetics},
  journal   = {Advances in Physics},
  volume    = {51},
  number    = {2},
  pages     = {481--587},
  year      = {2002},
  doi       = {10.1080/00018730110117433}
}

@book{onuki2002phase,
  author    = {Onuki, A.},
  title     = {Phase Transition Dynamics},
  publisher = {Cambridge University Press},
  address   = {Cambridge},
  year      = {2002}
}

@incollection{binder1991spinodal,
  author    = {Binder, K.},
  title     = {Spinodal Decomposition},
  booktitle = {Phase Transformations of Materials},
  editor    = {Cahn, R. W. and Haasen, P. and Kramer, E. J.},
  series    = {Materials Science and Technology},
  volume    = {5},
  pages     = {405--471},
  publisher = {VCH},
  address   = {Weinheim},
  year      = {1991}
}

@article{Jaiswal_2013,
doi = {10.1209/0295-5075/103/66003},
url = {https://doi.org/10.1209/0295-5075/103/66003},
year = {2013},
month = {oct},
publisher = {EDP Sciences, IOP Publishing and Società Italiana di Fisica},
volume = {103},
number = {6},
pages = {66003},
author = {Jaiswal, Prabhat K. and Puri, Sanjay and Binder, Kurt},
title = {Phase separation in thin films: Effect of temperature gradients},
journal = {Europhysics Letters}
}

@article{duhr2006,
  title = {Thermophoretic Depletion Follows Boltzmann Distribution},
  author = {Duhr, Stefan and Braun, Dieter},
  journal = {Phys. Rev. Lett.},
  volume = {96},
  issue = {16},
  pages = {168301},
  numpages = {4},
  year = {2006},
  month = {Apr},
  publisher = {American Physical Society},
  doi = {10.1103/PhysRevLett.96.168301},
  url = {https://link.aps.org/doi/10.1103/PhysRevLett.96.168301}
}

@article{lee2002,
author = {Lee, Kam-Wa D. and Chan, Philip K. and Feng, Xianshe},
title = {A Computational Study into Thermally Induced Phase Separation in Polymer Solutions under a Temperature Gradient},
journal = {Macromolecular Theory and Simulations},
volume = {11},
number = {9},
pages = {996-1005},
doi = {https://doi.org/10.1002/1521-3919(200211)11:9<996::AID-MATS996>3.0.CO;2-M},
url = {https://onlinelibrary.wiley.com/doi/abs/10.1002/1521-3919%28200211%2911%3A9%3C996%3A%3AAID-MATS996%3E3.0.CO%3B2-M},
eprint = {https://onlinelibrary.wiley.com/doi/pdf/10.1002/1521-3919%28200211%2911%3A9%3C996%3A%3AAID-MATS996%3E3.0.CO%3B2-M},
year = {2002}
}

@article{frenkel2015,
year = {2015},
volume = {143},
pages = {124104},
author = {Wirnsberger, P. and Frenkel, D. and Dellago, C.},
title = {An enhanced version of the heat
exchange algorithm with excellent energy
conservation properties},
journal = {The Journal of Chemical Physics},
}

@article{Gonnella_2008,
doi = {10.1088/1751-8113/41/10/105001},
url = {https://doi.org/10.1088/1751-8113/41/10/105001},
year = {2008},
month = {feb},
publisher = {},
volume = {41},
number = {10},
pages = {105001},
author = {Gonnella, G and Lamura, A and Piscitelli, A},
title = {Dynamics of binary mixtures in inhomogeneous temperatures},
journal = {Journal of Physics A: Mathematical and Theoretical}
}

@article{Hong_2010,
doi = {10.1088/0965-0393/18/2/025013},
url = {https://doi.org/10.1088/0965-0393/18/2/025013},
year = {2010},
month = {feb},
publisher = {},
volume = {18},
number = {2},
pages = {025013},
author = {Hong, Shujuan and Chan, Philip K},
title = {Simultaneous use of temperature and concentration gradients to control polymer solution morphology development during thermal-induced phase separation},
journal = {Modelling and Simulation in Materials Science and Engineering}
}

@article{voit2005,
  title = {Thermal Patterning of a Critical Polymer Blend},
  author = {Voit, A. and Krekhov, A. and Enge, W. and Kramer, L. and K\"ohler, W.},
  journal = {Phys. Rev. Lett.},
  volume = {94},
  issue = {21},
  pages = {214501},
  numpages = {4},
  year = {2005},
  month = {Jun},
  publisher = {American Physical Society},
  doi = {10.1103/PhysRevLett.94.214501},
  url = {https://link.aps.org/doi/10.1103/PhysRevLett.94.214501}
}

@article{kurita2017,
  title = {Control of pattern formation during phase separation initiated by a propagated trigger},
  author = {Kurita, Rei},
  journal = {Scientific Reports},
  volume = {7},
  issue = {1},
  pages = {6912},
  numpages = {4},
  year = {2017},
  month = {July},
  doi = {10.1038/s41598-017-07352-z},
  url = {https://doi.org/10.1038/s41598-017-07352-z}
}

@article{krekhov2009,
  title = {Formation of regular structures in the process of phase separation},
  author = {Krekhov, Alexei},
  journal = {Phys. Rev. E},
  volume = {79},
  issue = {3},
  pages = {035302},
  numpages = {4},
  year = {2009},
  month = {Mar},
  publisher = {American Physical Society},
  doi = {10.1103/PhysRevE.79.035302},
  url = {https://link.aps.org/doi/10.1103/PhysRevE.79.035302}
}

@article{mc-muglia-2012,
  title = {Dynamical and stationary critical behavior of the Ising ferromagnet in a thermal gradient},
  author = {Muglia, J. and Albano, E. V.},
  journal = {The European Physical Journal B},
  volume = {85},
  issue = {8},
  year = {2012}
}

@article{wave2009,
  title = {Traveling spatially periodic forcing of phase separation},
  author = {Weith, V. and Krekhov, A. and Zimmermann, W.},
  journal = {The European Physical Journal B},
  volume = {67},
  issue = {3},
  pages = {1434-6036},
  year = {2009},
  month = {Feb}
}

@article{LBpre2012,
  title = {Survey of morphologies formed in the wake of an enslaved phase-separation front in two dimensions},
  author = {Foard, E. M. and Wagner, A. J.},
  journal = {Phys. Rev. E},
  volume = {85},
  issue = {1},
  pages = {011501},
  numpages = {14},
  year = {2012},
  month = {Jan},
  publisher = {American Physical Society},
  doi = {10.1103/PhysRevE.85.011501},
  url = {https://link.aps.org/doi/10.1103/PhysRevE.85.011501}
}

@article{prl2006biro,
  title = {Phase Separation in the Wake of Moving Fronts},
  author = {Hantz, P\'eter and Bir\'o, Istv\'an},
  journal = {Phys. Rev. Lett.},
  volume = {96},
  issue = {8},
  pages = {088305},
  numpages = {4},
  year = {2006},
  month = {Mar},
  publisher = {American Physical Society},
  doi = {10.1103/PhysRevLett.96.088305},
  url = {https://link.aps.org/doi/10.1103/PhysRevLett.96.088305}
}

@article{furukawa-physica1992,
title = {Phase separation by directional quenching and morphological transition},
journal = {Physica A: Statistical Mechanics and its Applications},
volume = {180},
number = {1},
pages = {128-155},
year = {1992},
issn = {0378-4371},
doi = {https://doi.org/10.1016/0378-4371(92)90111-3},
url = {https://www.sciencedirect.com/science/article/pii/0378437192901113},
author = {Hiroshi Furukawa}
}

@article{foard-pre2009,
  title = {Enslaved phase-separation fronts in one-dimensional binary mixtures},
  author = {Foard, E. M. and Wagner, A. J.},
  journal = {Phys. Rev. E},
  volume = {79},
  issue = {5},
  pages = {056710},
  numpages = {15},
  year = {2009},
  month = {May},
  publisher = {American Physical Society},
  doi = {10.1103/PhysRevE.79.056710},
  url = {https://link.aps.org/doi/10.1103/PhysRevE.79.056710}
}

@article{gonnella-pre2010,
  title = {Phase separation of binary fluids with dynamic temperature},
  author = {Gonnella, G. and Lamura, A. and Piscitelli, A. and Tiribocchi, A.},
  journal = {Phys. Rev. E},
  volume = {82},
  issue = {4},
  pages = {046302},
  numpages = {8},
  year = {2010},
  month = {Oct},
  publisher = {American Physical Society},
  doi = {10.1103/PhysRevE.82.046302},
  url = {https://link.aps.org/doi/10.1103/PhysRevE.82.046302}
}

@article{liang-prl2013,
  title = {Liquid Phase Stability Under an Extreme Temperature Gradient},
  author = {Liang, Zhi and Sasikumar, Kiran and Keblinski, Pawel},
  journal = {Phys. Rev. Lett.},
  volume = {111},
  issue = {22},
  pages = {225701},
  numpages = {5},
  year = {2013},
  month = {Nov},
  publisher = {American Physical Society},
  doi = {10.1103/PhysRevLett.111.225701},
  url = {https://link.aps.org/doi/10.1103/PhysRevLett.111.225701}
}

@article{zimmermann2022,
author = {Zimmermann, Nils E. R. and Guevara-Carrion, Gabriela and Vrabec, Jadran and Hansen, Niels},
title = {Predicting and Rationalizing the Soret Coefficient of Binary Lennard-Jones Mixtures in the Liquid State},
journal = {Advanced Theory and Simulations},
volume = {5},
number = {11},
pages = {2200311},
doi = {https://doi.org/10.1002/adts.202200311},
url = {https://advanced.onlinelibrary.wiley.com/doi/abs/10.1002/adts.202200311},
eprint = {https://advanced.onlinelibrary.wiley.com/doi/pdf/10.1002/adts.202200311},
year = {2022}
}

@book{crctemp,
  author    = {ME. Fisher},
  title     = {Critical phenomena / edited by M. S. Green},
  publisher = {Academic Press},
  address   = {London},
  year      = {1971}
}

@book{demixing,
  author    = {H.E. Stanley},
  title     = {Introduction to Phase Transitions and Critical Phenom-
ena},
  publisher = {Oxford University Press},
  address   = {Oxford},
  year      = {1971}
}

@book{CHC,
  author    = {S. Puri and V. Wadhawan (eds.)},
  title     = {Kinetics of Phase transitions},
  publisher = {CRC Press},
  address   = {Boca Raton},
  year      = {2009}
}

@book{func,
  author    = {M. Plischke and B. Bergersen},
  title     = {Equilibrium Statictical Mechanics},
  publisher = {World Scientific},
  address   = {Singapore},
  year      = {2006}
}

@article{Gomezprl2016,
  title = {Dynamics of Self-Propelled Janus Particles in Viscoelastic Fluids},
  author = {Gomez-Solano, Juan Ruben and Blokhuis, Alex and Bechinger, Clemens},
  journal = {Phys. Rev. Lett.},
  volume = {116},
  issue = {13},
  pages = {138301},
  numpages = {5},
  year = {2016},
  month = {Mar},
  publisher = {American Physical Society},
  doi = {10.1103/PhysRevLett.116.138301},
  url = {https://link.aps.org/doi/10.1103/PhysRevLett.116.138301}
}

@article{felix2024-jcp,
    author = {Höfling, F. and Dietrich, S.},
    title = {Structure of liquid–vapor interfaces: Perspectives from liquid state theory, large-scale simulations, and potential grazing-incidence x-ray diffraction},
    journal = {The Journal of Chemical Physics},
    volume = {160},
    number = {10},
    pages = {104107},
    year = {2024},
    month = {03},
    issn = {0021-9606},
    doi = {10.1063/5.0186955},
    url = {https://doi.org/10.1063/5.0186955},
    eprint = {https://pubs.aip.org/aip/jcp/article-pdf/doi/10.1063/5.0186955/19724417/104107_1_5.0186955.pdf},
}

@article{fayaz2022,
author = {Fayaz-Torshizi, Maziar and Xu, Weilun and Vella, Joseph R. and Marshall, Bennett D. and Ravikovitch, Peter I. and M{\"u}ller, Erich A.},
title = {Use of Boundary-Driven Nonequilibrium Molecular Dynamics for Determining Transport Diffusivities of Multicomponent Mixtures in Nanoporous Materials},
journal = {The Journal of Physical Chemistry B},
volume = {126},
number = {5},
pages = {1085-1100},
year = {2022},
doi = {10.1021/acs.jpcb.1c09159},
    note ={PMID: 35104134},

URL = {https://doi.org/10.1021/acs.jpcb.1c09159}
}

@article{sahu_2025,
doi = {10.1088/1402-4896/adc5b2},
url = {https://doi.org/10.1088/1402-4896/adc5b2},
year = {2025},
month = {apr},
publisher = {IOP Publishing},
volume = {100},
number = {5},
pages = {055931},
author = {Sahu, Geetika and Chakraborty, Chanchal and Roy, Subhadeep and Banerjee, Souri},
title = {Non-monotonic growth of MoS2 quantum dots observed through fractal analysis: role of precursor concentration on the overall growth dynamics},
journal = {Physica Scripta}
}
\end{document}